\documentclass[9pt,twocolumn,twoside]{opticajnl}
\journal{opticajournal} 

\setboolean{shortarticle}{true}

\usepackage{lineno}
\usepackage{bm}

\title{Phase-matched locally chiral light for global control of chiral light-matter interaction}

\author[1,*]{Chong Ye}
\author[1]{Yifan Sun}
\author[2]{Libin Fu}
\author[1]{Xiangdong Zhang}

\affil[1]{Beijing Key Laboratory of Nanophotonics and Ultrafine Optoelectronic Systems, School of Physics, Beijing Institute of Technology, Beijing 100081, China}
\affil[2]{Graduate School of China Academy of Engineering Physics, Beijing 100193, China}

\affil[*]{yechong@bit.edu.cn}

\begin{abstract}
Locally chiral light is an emerging tool for probing and controlling molecular chirality. It can generate large and freely adjustable enantioselectivities in purely electric-dipole effects, offering its major advantages over traditional chiral light. However, the existing types of locally chiral light are phase-mismatched, and thus the global efficiencies are greatly reduced compared with the maximum single-point efficiencies or even vanish. Here, we propose a scheme to generate phase-matched locally chiral light. To confirm this advantage, we numerically show the robust highly efficient global control of enantiospecific electronic state transfer of methyloxirane at nanoseconds. Our work potentially constitutes the starting point for developing more efficient chiroptical techniques for the studies of chiral molecules.
\end{abstract}

\setboolean{displaycopyright}{false} 

\begin{document}

\maketitle

Molecular chirality plays a central role in chemical reactions, biological activities, drug industries, and life's chemistry, which makes probing and controlling molecular chirality vitally important and challenging tasks across natural science
. 
Using single circularly light (i.e., traditional chiral light) to probe~\cite{berova2000circular,barron2009molecular,busch2011chiral,nafie2011vibrational} molecular chirality began with the first discovery of molecular chirality. It was also proposed to control enantiomeric excess in an inner state~\cite{salam1998enantiomeric}, i.e., enantiospecific state transfer. The enantioselectivities generated by the traditional chiral light come from the interference between electric- and magnetic-dipole light-matter interactions. The relative strength between them is almost a fixed parameter at a ratio of ($\simeq1/137$)~\cite{salam1998enantiomeric}, so the generated enantioselectivities are usually tiny and not freely adjustable. 

Locally chiral light (LCL)~\cite{ayuso2019synthetic,ayuso2022ultrafast,neufeld2021strong,neufeld2021detecting,ayuso2021ultrafast,ayuso2021enantio,khokhlova2022enantiosensitive} is an emerging chiral light source that can generate large and freely adjustable enantioselectivities in purely electric-dipole effects. LCLs are multi-chrome, composed of carrier waves with well-designed polarizations and frequencies. While the traditional chiral light's degree of chirality (DOC) at a single point in space is determined by the electric and magnetic fields at that point and nearby~\cite{tang2010optical}, LCLs' DOC at a single point only depends on the electric field at that point (or equivalently the local electric field). {Specifically}, at each fixed point in space, the tip of the electric field vector draws a chiral 3D Lissajous curve when it evolves in time. Beyond probing and controlling chiral molecules, the LCLs were also proposed to imprint chirality on achiral matter~\cite{mayer2022imprinting}, offering new opportunities to realize laser-driven achiral–chiral phase transitions in matter~\cite{thirunamachandran1977laser,wismer2017ultrafast,owens2018climbing}.

With the benefit of hindsight, the studies of all-electric-dipole chiroptical techniques by using LCLs began almost two decades ago in the framework of the cyclic three-level model~\cite{kral2001cyclic,kral2003two,ye2018real,leibscher2019principles}, where the driving electromagnetic fields are tricolor LCLs. The three carrier waves couple with chiral molecules in the one-photon (near-)resonance, {and thus the dynamics of chiral molecules can be described by the cyclic three-level model}~\cite{shapiro2000coherently,li2007generalized,vitanov2019highly,ye2021entanglement,cai2022enantiodetection,leibscher2022full,ye2023single}. Experimentally, the tricolor LCLs were used in the well-demonstrated microwave techniques of enantiodetection~\cite{patterson2013enantiomer,patterson2013sensitive,patterson2014new,shubert2014identifying,lobsiger2015molecular,shubert2016chiral} and enantio-specific state transfer~\cite{eibenberger2017enantiomer,perez2017coherent,lee2022quantitative}. The equivalent chiroptical techniques in the UV-IR region~\cite{kral2001cyclic} are long pursued but never realized. One key problem is that tricolor LCLs are phase-mismatched~\cite{lehmann2018theory}, such that their DOCs and the generated enantioselectivities change periodically in space and even vanish by integrating over the whole spatial period (or equivalently globally achiral). This problem becomes grave in the UV-IR region because of the short wavelengths~\cite{ayuso2022ultrafast}. 

{To solve this problem, chirality polarized light (CPL)~\cite{ayuso2021enantio} and locally-and-globally chiral light (LGCL)~\cite{ayuso2019synthetic} were proposed. CPL is globally achiral but has the polarization of chirality, just like the neutral and polarized one-dimensional medium of alternating negative and positive charges. This property makes the emission directions highly enantioselective for CPL.} LGCL is more efficient because it is 
globally chiral. However, they are still phase-mismatched, such that the enantiospecific electronic state transfer in the UV-IR region is still out of reach. Moreover, CPL and LGCL worked in the strong-field region, {which is usually accompanied by the destruction of the molecule or the change of the molecular conformer (not chirality).}

In this Letter, we propose a scheme to generate a novel type of LCL without the phase-mismatch problem. Thus, we name it the phase-matched LCL. {This feature offers our scheme advantages over existing LCLs in global control of chiral light-molecule interactions.} To illustrate this, we use a four-color example of our scheme and numerically show its ability in the robust highly efficient global control of enantiospecific electronic state transfer of methyloxirane at nanoseconds in the UV-IR region.

\begin{figure}
	\centering
	\includegraphics[width=0.99\columnwidth]{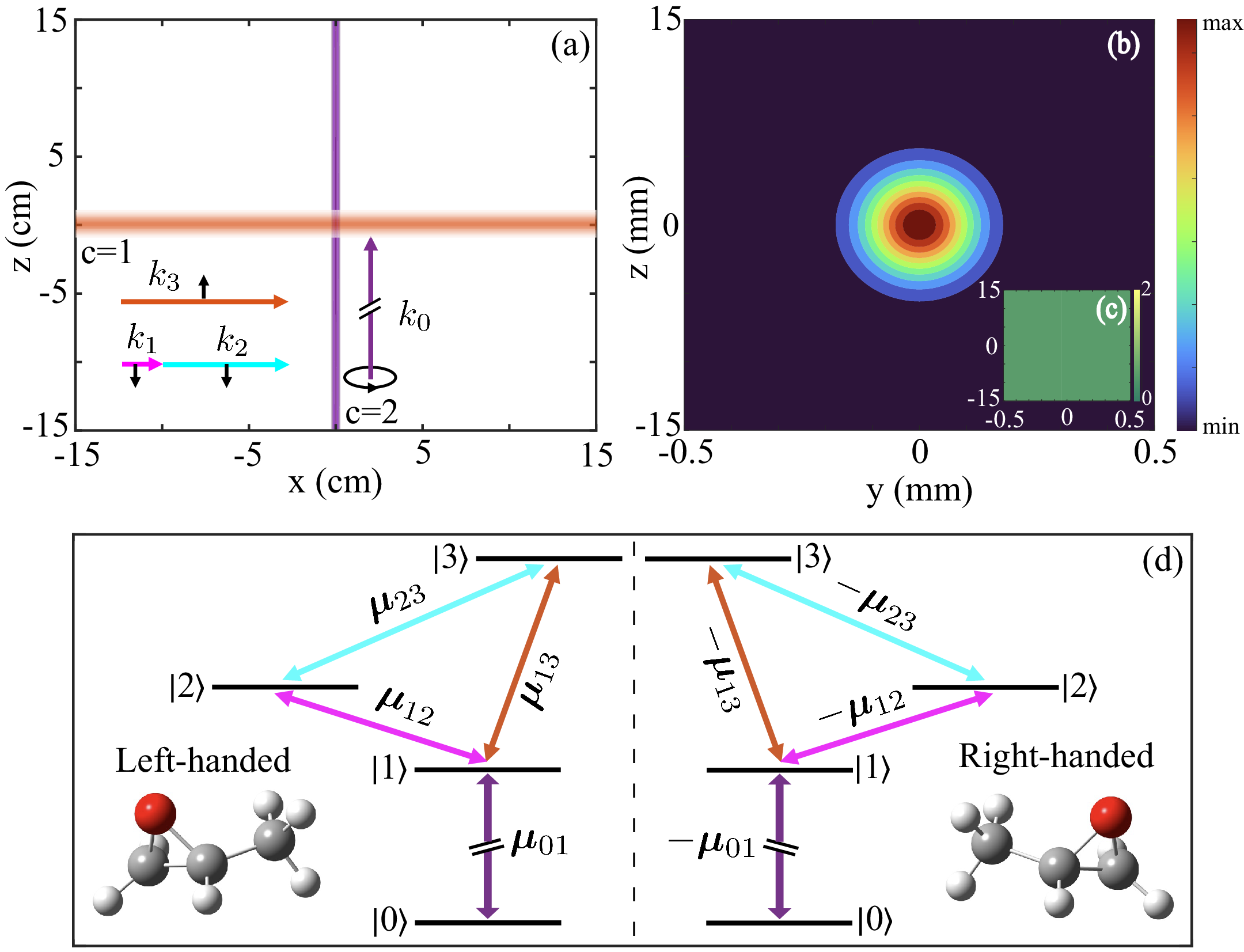}\\
	\caption{(a) Four-color phase-matched LCL: the $c=1$ sub-beam is a $z$-polarized three-color field propagating in the $x$-direction, and the $c=2$ sub-beam is a one-color circularly polarized field propagating in the $z$-direction. The carrier waves are Gaussian beams with waists of $w_{0,0}=10^3\lambda_0$ and $w_{0,1}=w_{0,2}=w_{0,3}=10^3\lambda_1$. The wavelengths $\lambda_0$ and $\lambda_1$ are chosen according to the example of methyloxirane in Fig.\,\ref{FigPS}. The lowest-order CCFs [$h^{(5)}$] in the $y-z$ plane at $z=0$ clearly show its phase-matched feature: (b) the corresponding amplitudes (in arbitrary unit) and (c) phases (in the unit of $\pi$). (d) Scheme of enantioselective light-molecule interaction for our four-color LCL in the one-photon (near-)resonance. The enantioselectivity is reflected in the transition electric dipoles with $\bm{\mu}^{L}_{ij}=-\bm{\mu}^{R}_{ij}=\bm{\mu}_{ij}$.}\label{FigA}
\end{figure}

Our scheme consists of two non-collinear sub-beams. The first sub-beam [labeled with $c=1$ in Fig.\ref{FigA}\,(a)] is linearly polarized and multi-chrome. The frequencies of carrier waves therein are in the cyclic odd-photon resonance. For the case of a four-color example [see Fig.\,\ref{FigA}\,(a)], such a condition is $\omega_1+\omega_2-\omega_{3}=0$, yielding $\bm{k}_1+\bm{k}_2-\bm{k}_3=0$ [see the left lower corner of Fig.\,\ref{FigA}\,(a)]. The second sub-beam [labeled with $c=2$ in Fig.\ref{FigA}\,(a)] is monochrome and exhibits circular polarization in the plane perpendicular to the polarization of the first sub-beam. 

\begin{figure}
	\centering
	\includegraphics[width=0.99\columnwidth]{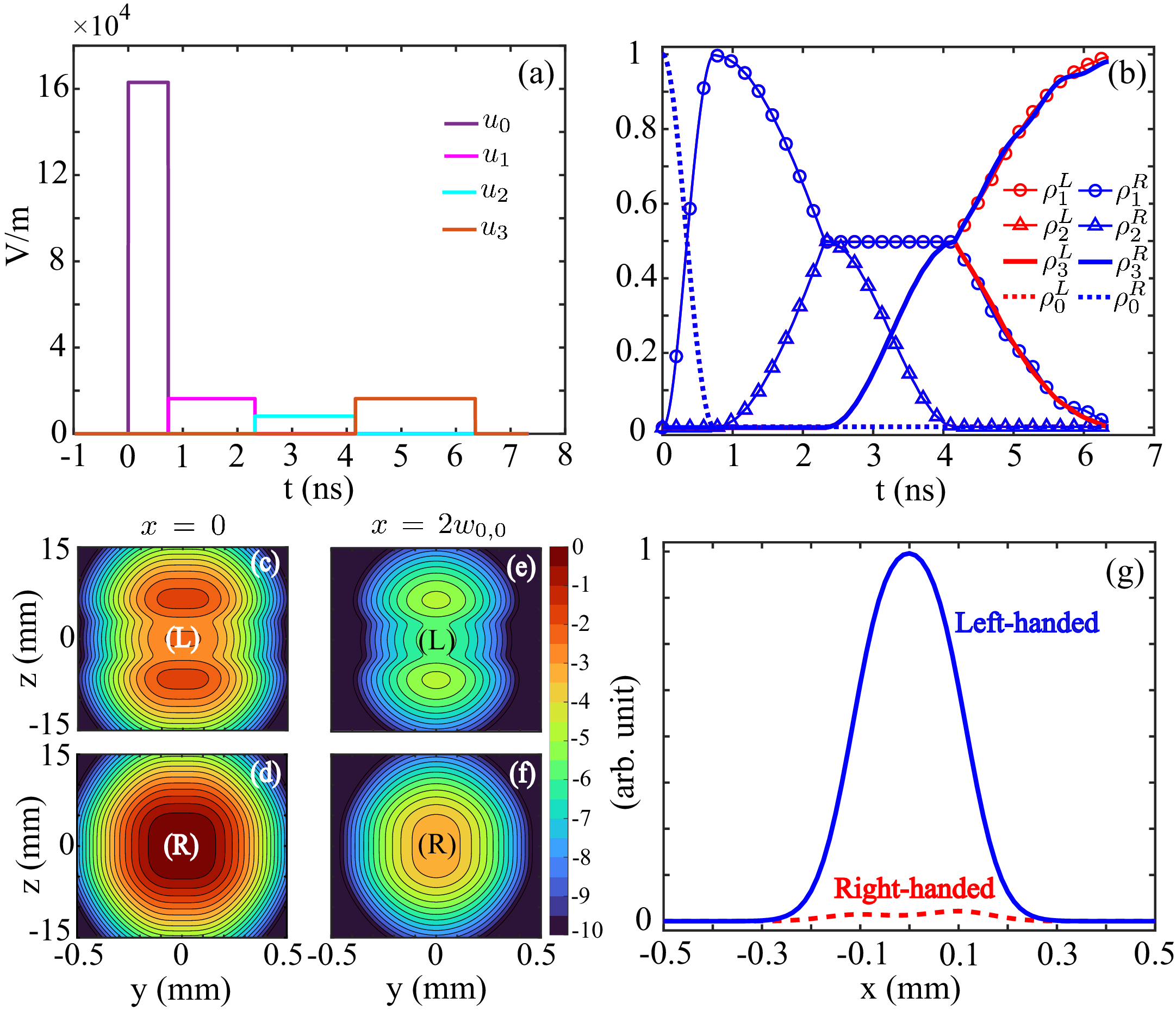}\\
	\caption{Global control of enantiospecific electronic state transfer of methyloxirane by using our four-color phase-matched LCL. (a) Temporal envelopes of the applied fields at $r=0$. (b) Time evolution of the population in each working electronic state for molecules at $r=0$, yielding the enantiomeric excess of $-99.27\%$ at $r=0$. (c-f) $\log_{10}(\rho^{L}_3)$ and $\log_{10}(\rho^{R}_3)$ after the four pulses in the $y-z$ plane for $x=0$ and $x=2w_{0,0}\simeq 0.32$\,mm. (g) $P^{Q}_{2D}(x)\equiv\int dydz\rho^{Q}_3(\bm{r})$ with $Q(=L,R)$ denoting the handedness of the chiral molecules. The global enantiomeric excess is $\varepsilon_{\mathrm{global}}\simeq-93.74\%$.}\label{FigPS}
\end{figure}

The overall field of the four-color example is
\begin{align}\label{OFE}
\bm{E}(\bm{r},t)=&\sum^{N}_{m=1} \frac{{E}_{m}}{2}\bm{e}_z e^{-i(\omega_m t- k_m x)}
+\frac{{E}_{0}}{2}e^{-i(\omega_0 t-k_0 z)}\bm{e}_{\sigma}+c.c.,
\end{align}
where $\sigma(=\pm1)$ denotes the handedness of the circularly polarized field and the corresponding polarization vector is $\bm{e}_{\sigma}=(\sigma\bm{e}_x+i\bm{e}_y)/\sqrt{2}$. The amplitudes and phases of ${E}_{i}$ are $\mathcal{E}_i$ and $\phi_i$. For the four-color example of Fig.\,\ref{FigA}, we have $N=3$. We choose the carrier waves to be Gaussian beams [see Fig.\,\ref{FigA}\,(a)]. Their amplitudes and phases are  $\mathcal{E}_{i}=u_{i}w_{0,i}e^{-r^2_{\perp_i}/w^2_{i}}/w_{i} $ and $\phi_i=k_{i}r^2_{\perp_i}/[4r^2_{\parallel_i}+(k_{i}w^2_{0,i})^2]-\arctan[2 r_{\parallel_i}/(k_{i} w^2_{0,i})]+\varphi_i$ with $w_i=w_{0,i}\sqrt{1+[2 r_{\perp_i}/(k_i w^2_{0,i})]^2}$, where $r_{\parallel_i}$ and $r_{\perp_i}$ are projections of the position vectors parallel and vertical to the corresponding propagation directions. Their waists are chosen as $w_{0,0}=10^3\lambda_0$ and $w_{0,1}=w_{0,2}=w_{0,3}=10^3\lambda_1$. The initial phases are $\varphi_i$. Here, $\lambda_0$ and $\lambda_1$ are wavelengths of $E_0$ and $E_{1}$, whose values are chosen according to the example of methyloxirane in Fig.\,\ref{FigPS}.

The DOC of locally chiral light at a single point in space is characterized by the $n$th order chiral correlation functions (CCFs)~\cite{ayuso2022ultrafast} defined as $h^{(n)}(\bar{\omega}_0,...,\bar{\omega}_{n-1})\equiv\{\bm{F}(\bar{\omega}_0)\cdot [\bm{F}(\bar{\omega}_1)\times \bm{F}(\bar{\omega}_2)]\}
 [\bm{F}(\bar{\omega}_3)\cdot \bm{F}(\bar{\omega}_{4})]...[\bm{F}(\bar{\omega}_{n-2})\cdot \bm{F}(\bar{\omega}_{n-1})]$ with the cyclic odd-photon resonance condition of $\bar{\omega}_i$. Here, $\bm{F}(\bar{\omega}_i)$ is a Fourier transform of the electric field vector at the angular frequency $\bar{\omega}_i$. We can prove that all the nonzero CCFs of the overall field in Eq.\,(\ref{OFE}) are irrelevant to the spatially oscillated phase of each carrier wave $\bm{k}_i\cdot \bm{r}$. That is to say, the proposed LCLs in Eq.\,(\ref{OFE}) are phase-matched. For more details on the proof, see Sec.\,I\,A of Supplement. 
 
 Specifically, we show the amplitude [Fig.\,\ref{FigA}\,(b)] and phase [Fig.\,\ref{FigA}\,(c)] of lowest-order CCFs [$h^{(5)}$] in the $y-z$ plane at $z=0$ for the four-color example. The results clearly show that the four-color example is phase-matched. In contrast, the nonzero CCFs of the tricolor LCL and LGCL are functions of $\bm{k}_i\cdot \bm{r}$ (see Sec.\,I\,C and Sec.\,I\,D of Supplement), i.e., they are phase-mismatched. We note that Eq.\,(\ref{OFE}) provides a general scheme to form phase-matched LCLs, including the two-color phase-matched LCL (see Sec.\,I\,B of Supplement). 

For the four-color example, {we are interested in the case of the one-photon (near-)resonance.} {Because the residual electronic transitions are coupled with the overall field off-resonantly, they are negligible in the weak-field region, yielding the simplified working model as shown in Fig.\,\ref{FigA}\,(d).}
The molecules also have vibrational and rotational degrees of freedom. We deal with them in the Born-Oppenheimer approximation. According to the Franck–Condon principle~\cite{atkins2011molecular}, which says that an electronic transition occurs within a stationary nuclear framework, the vibrational degrees of freedom can be assumed frozen, 
i.e., the electronic transitions are vertical transitions. {In our discussions, we are interested in the gaseous phase chiral molecules. For them,} the rotational degree of freedom is described by 
$\hat{H}_{\mathrm{rot}}=\hbar (A\hat{J}^2_a+B\hat{J}^2_b+C\hat{J}^2_c)$~\cite{zare1988angular}.
$\hat{J}_{a,b,c}$ are angular momentum operators along
the principal axes of the moment of inertia, respectively. $A$, $B$, and $C$ are the corresponding rotational 
constants. Because the nuclear locations are preserved in electronic vertical transitions, the rotational 
constants are preserved in our consideration. 

In the rotating-wave approximation, the concrete Hamiltonian of the chiral light-molecule interaction is
 \begin{align}\label{HM}
\hat{H}=&\sum_{\alpha,\beta}[\frac{1}{2}(\Omega^{\alpha\beta}_{10}\hat{s}_{10}+\sum^{3}_{i>j=1}\Omega^{\alpha\beta}_{ij}\hat{s}_{ij})\hat{\sigma}_{\alpha\beta}+h.c.]+\sum_{i\ne 1}\Delta_{i1}\hat{s}_{ii}+\hat{H}_{\mathrm{rot}}.
\end{align}
The detunings are $\Delta_{01}=\omega_0-(v_1-v_0)$, $\Delta_{21}=v_2-v_1-\omega_1$, $\Delta_{31}=v_3-v_1-\omega_3$, and $\Delta_{32}=v_3-v_2-\omega_2$. {$\hat{\sigma}_{\alpha\beta}\equiv|{\alpha}\rangle\langle {\beta}|$ are operators in the rotational degree of freedom with $|\alpha\rangle\equiv|J^{\alpha}_{K^{\alpha}_a,K^{\alpha}_c,M^{\alpha}}\rangle$ being the asymmetric-top rotational eigenstates~\cite{zare1988angular}. We note that $\sum_{\alpha}$ means that the summation is taken over all $J^{\alpha}$, $K^{\alpha}_a$, $K^{\alpha}_c$, and $M^{\alpha}$.} $\hat{s}_{ij}$ are operators in the electronic degree of freedom. Because the transition electric dipoles change signs with enantiomers [see Fig.\ref{FigA}\,(a)], the interaction Hamiltonian~(\ref{HM}) is enantioselective as indicated by $\Omega^{\alpha\beta}_{ij,L}=-\Omega^{\alpha\beta}_{ij,R}$ with the subscripts ``$L$" and ``$R$" denoting the chirality. We note that the operators 
$\hat{s}_{ij}\equiv|\tilde{i}\rangle\langle \tilde{j}|$ are defined in the new basis with $|\tilde{0}\rangle=e^{-\mathrm{i}[(v_1-\omega_0)t+\bm{k}_0\cdot\bm{r}]}|0\rangle$, $|\tilde{1}\rangle=e^{-\mathrm{i}v_1 t}|1\rangle$, $|\tilde{2}\rangle=e^{-\mathrm{i}[(v_1+\omega_1)t-\bm{k}_1\cdot \bm{r}]}|2\rangle$, and $|\tilde{3}\rangle=e^{-\mathrm{i}[(v_1+\omega_3)t-\bm{k}_3\cdot \bm{r}]}|3\rangle$. Then, the spatially and temporally oscillated phases $(\bm{k}_i\cdot \bm{r}-\omega_i t)$ are transferred to $\Omega^{\alpha\beta}_{32}e^{i(\delta \omega t-\delta\bm{k}\cdot \bm{r})}$ with $\delta\omega\equiv\omega_3-\omega_2-\omega_1=0$ and $\delta\bm{k}\equiv\bm{k}_3-\bm{k}_2-\bm{k}_1=0$. This clearly indicates that the Hamiltonian~(\ref{HM}) is phase-matched, such that the molecules at different positions evolve identically in the plane-wave limit.

To illustrate the advantages of our four-color phase-matched LCL, we use methyloxirane as an example, which is a typical chiral molecule [see Fig.\,\ref{FigA}\,(d)]. The working electronic states $|0\rangle$, $|1\rangle$, $|2\rangle$, and $|3\rangle$ are chosen as the electronic ground state, the $3p_y$-state, the $3p_x$-state, and the $3d_{z^2-x^2}$-state~\cite{khokhlova2022enantiosensitive}. The carrier waves with angular frequencies $\omega_0$, $\omega_1$, $\omega_2$, and $\omega_3$ are chosen to resonantly couple with transitions $|0\rangle|0_{0,0}\rangle\leftrightarrow|1\rangle|1_{0,1}\rangle$, $|1\rangle|1_{0,1}\rangle\leftrightarrow|2\rangle|2_{1,1}\rangle$, $|2\rangle|2_{1,1}\rangle\leftrightarrow|3\rangle|2_{0,2}\rangle$, and $|1\rangle|1_{0,1}\rangle\leftrightarrow|3\rangle|2_{0,2}\rangle$. The corresponding wavelengths are $\lambda_0\simeq162.2633$\,nm, $\lambda_1\simeq6241.4404$\,nm, $\lambda_2\simeq1541.6517$\,nm, and $\lambda_3\simeq2047.3531$\,nm~\cite{khokhlova2022enantiosensitive}. The spatial profiles of carrier waves are shown in Fig.\,\ref{FigA}\,(a).

To achieve enantiomeric excess in the state $|3\rangle$, we take the temporally square pulses, whose amplitudes at $r=0$ evolve as shown in Fig.\,\ref{FigPS}\,(a). 
To justify the validity of using the classical description of light in our discussions, we give the average photon number of the pulses, which are about $6.2\times 10^{15}$, $7.7\times10^{18}$, $8.7\times10^{17}$, and $1.2\times10^{18}$. They are large enough to make the photon-number phase uncertainty negligible~\cite{gerry2005introductory}. The two enantiomers are assumed initially in the state $|0\rangle|0_{0,0,0}\rangle$. This is a good approximation at the rotational temperate $T_{\mathrm{rot}}\lesssim 10$\,mK because the relative populations in the first excited rotational state with respect to the ground rotational state is $\lesssim7\times 10^{-6}$.

The evolution of the population in each electronic state at $r=0$ is shown in Fig.\,\ref{FigPS}\,(b), {which is given by summing over its rotational sub-states.} We note that the light-molecule interactions in sufficiently high $J$ are negligible because these transitions are off-resonantly coupled with light. {That is to say, although the number of rotational levels is infinite, the calculations can be done in the subspace with $J\le J_c$. We numerically find that the calculations converge at $J_c=3$ (see Sec.\,II\,A of Supplement).}
{When all pulses end}, while almost all the population of the right-handed molecule is transferred to the electronic state $|3\rangle$, most of the left-handed molecule's population is transferred to the electronic state $|2\rangle$. We focus on the enantiomeric excess in the electronic state $|3\rangle$ defined as $\varepsilon\equiv ({\rho^{L}_{3}-\rho^{R}_{3}})/({\rho^{L}_{3}+\rho^{R}_{3}})$, where $\rho^{L}_{3}$ and $\rho^{R}_{3}$ are the final populations in the electronic state $|3\rangle$ for the two enantiomers. The achieved enantiomeric excess corresponding to Fig.\,\ref{FigPS}\,(b) is about $\varepsilon\simeq 99.27\%$. 

{Usually, the $M$-degeneracy problem and the broad frequency features of applied fields significantly reduce enantioselectivity. In our calculations, without their influences, the applied pulses will form perfect ``$\pi-{\pi}/{2}-\pi-\pi/2$" pulses, yielding the final enantiomeric excess of $100\%$. Here, we carefully design the polarization and the center frequency of each field, these problems are largely resolved at nanoseconds, and thus these pulses did not deviate too much from perfect $\pi$ and $\pi/2$ pulses. Then, the final high enantiomeric excess is considerably high.}

In Fig.\,\ref{FigPS}\,(c-f), we give $\log_{10}(\rho^{L}_3)$ and $\log_{10}(\rho^{R}_3)$ in the $y-z$ plane at $x=0$ and $x=2 w_{0,0}$. In the case of $x=0$, $\rho^{L}_3$ [see Fig.\,\ref{FigPS}\,(c)] are much smaller than $\rho^{R}_3$ [see Fig.\,\ref{FigPS}\,(d)] near the origin of the plane. When the molecules are away from the origin of the plane, $\rho^{L}_3$ and $\rho^{R}_3$ decrease to $zero$. Because the waist of the $\omega_0$ field is much smaller than those of others, the decreasing rate in the $y$-direction is much larger than that in the $z$-direction. In the case of $x=2 w_{0,0}$ [see Fig.\,\ref{FigPS}\,(e) and Fig.\,\ref{FigPS}\,(f)], similar phenomena happen in the $y-z$ plane.  
 
Further, we sum $\rho^{L}_3$ and $\rho^{R}_3$ over the $y-z$ plane at different $x$, yielding $P^{Q}_{2D}(x)\equiv\int dydz\rho^{Q}_3(\bm{r})$. The results in Fig.\,\ref{FigPS}\,(g) show that $P^{R}_{2D}$ and $P^{L}_{2D}$ decrease when the plane is moved away from $x=0$ in the $x$-direction. To evaluate global efficiency, we define the global enantiomeric excess 
\begin{align}
\varepsilon_{\mathrm{global}}\equiv\frac{P^L_{3D}-P^R_{3D}}{P^L_{3D}+P^R_{3D}}
\end{align}
with $P^{Q}_{3D}\equiv\int d\bm{r}^3\rho^{Q}_3(\bm{r})$.
The corresponding global enantiomeric excess is $\varepsilon_{\mathrm{global}}\simeq-93.74\%$. This shows considerably high efficiency in the global control of chiral light-molecule interactions. For the tricolor LCLs, $\varepsilon_{\mathrm{global}}$ quickly decreases to $zero$ when the size of the sample becomes comparable to the wavelengths (see Sec.\,II\,B of Supplement). Thus, our four-color LCLs show a considerable advantage in global control of chiral light-molecule interaction. {At typical supersonic beam temperature ($T_{\mathrm{rot}}=1$\,K)~\cite{lee2022quantitative} and with the same four-color LCLs, the enantiomeric excess at the origin is about $93\%$ and the global enantiomeric excess is about $91\%$ (see Sec.\,II\,A of Supplement).} 

 \begin{figure}
	\centering
	\includegraphics[width=0.99\columnwidth]{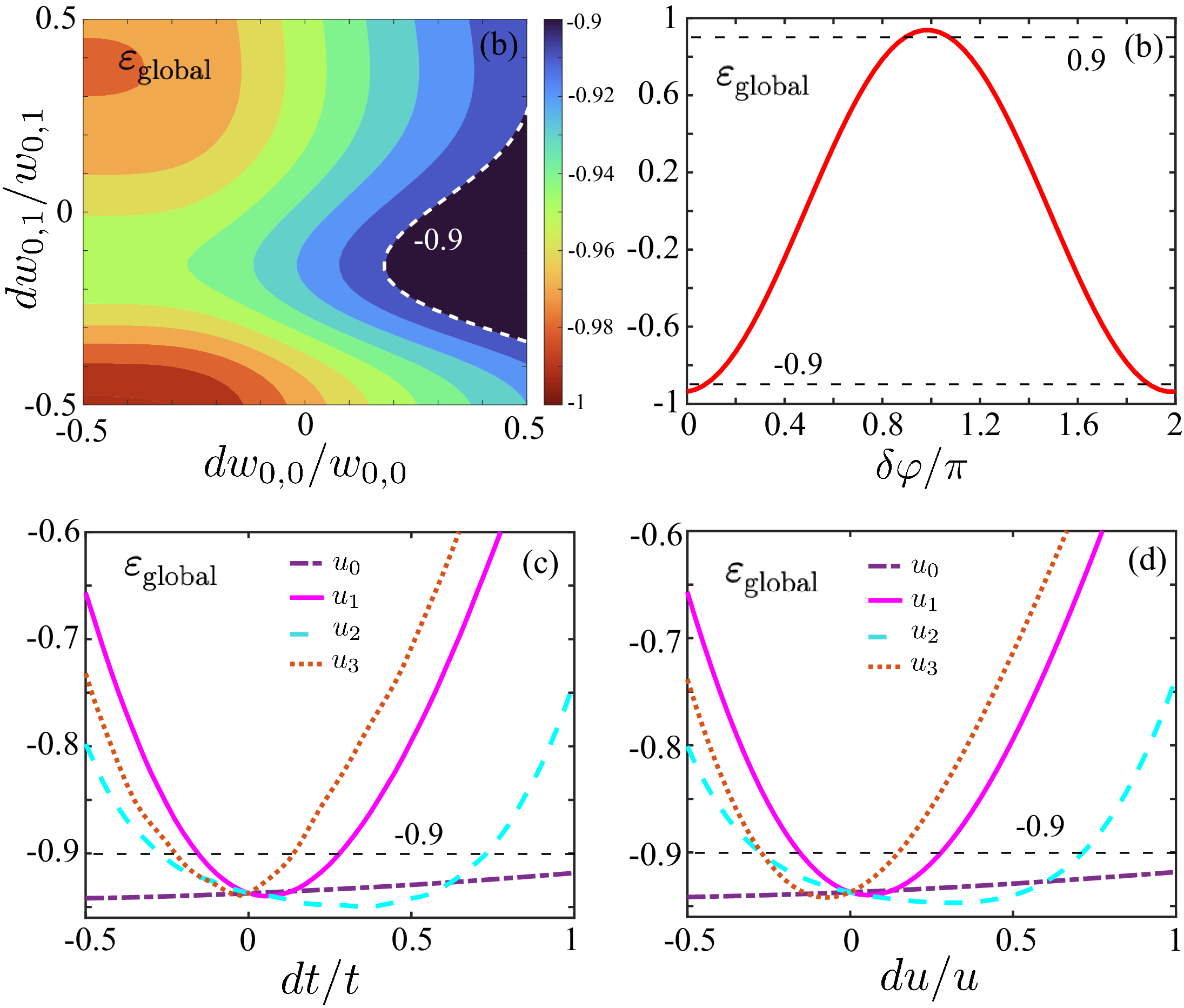}\\
	\caption{Robustness of highly efficient global control of enantiomeric excess ($|\varepsilon_{\mathrm{global}}|\ge 0.9$)  against relative control errors with respect to their corresponding values in Fig.\,\ref{FigPS}: (a) the waists $w_{0,0}$ and $w_{0,1}$ ($w_{0,2}=w_{0,3}=w_{0,1}$), (b) the phases $\delta\varphi$, (c) the duration of each pulse, and (d) the field amplitude of each pulse.} \label{Fig3}
\end{figure}

We also explore the robustness against control errors at $T_{\mathrm{rot}}\le 10$\,mK (see Fig.\,\ref{Fig3}). {When the relative variations of $w_{0,0}$ and $w_{0,1}$ are limited to $\pm 50\%$, the worst result is $\varepsilon_{\mathrm{global}}\simeq -87\%$ [see Fig.\,\ref{Fig3}\,(a)]. That is to say, the high global efficiency is robust against the variations of the waists of the beams. As shown in Fig.\,\ref{Fig3}\,(b), the variation of $\delta\varphi\equiv\varphi_3-\varphi_2-\varphi_1$ can affect the enantiomeric excess. The highly efficient global control ($|\varepsilon_{\mathrm{global}}|\ge 90\%$) can be achieved around $\delta\varphi=0$ and $\delta\varphi=\pi$. We note that the variation of $\phi_0$ will not {affect} the enantiomeric excess. As shown in Fig.\,\ref{Fig3}\,(c), the high global efficiency is robust against the relative variations of the pulses' duration ($dt/t$). For $E_0$, $E_1$, $E_3$, and $E_2$ pulses, $|\varepsilon_{\mathrm{global}}|\ge 90\%$ can be obtained with the relative variations about $-50\%\sim 100\%$, $-15\%\sim 30\%$, $-25\%\sim 75\%$, and $-25\%\sim 15\%$, respectively. Each line in Fig.\,\ref{Fig3}\,(c) is obtained by changing the duration of the corresponding pulse around its value in Fig.\,\ref{FigPS} and fixing the other pulses' duration to their values in Fig.\,\ref{FigPS}. Correspondingly, the pulses' duration for the four carrier waves can be tuned in the regions about $0.37\sim1.10$\,ns, $1.35\sim2.06$\,ns, $1.38\sim3.22$\,ns, and $1.65\sim2.52$\,ns, respectively. In Fig.\,\ref{Fig3}\,(d), we show $\varepsilon_{\mathrm{global}}$ as the function of the variation of each pulse's amplitude around the corresponding value in Fig.\,\ref{FigPS}, which also show considerable robustness.}


We believe our scheme provides a promising chiral light source in the UV-IR region. It purely depends on strong electric-dipole effects and is phase-matched. {Our results clearly show its advantage in global control of enantiospecific state transfer. It constitutes the starting point for the applications of other UV-IR spectroscopic techniques in the studies of chiral molecules.}

{We note that there are other factors that limit the enantiospecific state transfer in the UV-IR region, including the involvement of multiple rotational levels due to the broad-frequency features of the applied fields and the short lifetime of electronic excited states. These problems exist for all the chiral light source, and should be addressed in further research.}


\begin{backmatter}

\bmsection{Acknowledgments} 
This work is supported by the National Natural Science Foundation of China (No.\,12105011, No.\,12234004, No.\,12088101, No. U2330401, and No.\,11904022).


\bmsection{Disclosures} The authors declare no conflicts of interest.

\bmsection{Data availability} No data were generated or analyzed in the presented research.

\bmsection{Supplemental document}
See Supplement for supporting content. 

\end{backmatter}

\bibliography{ycref}

\bibliographyfullrefs{ycref}



\end{document}


\maketitle

\section{Chiral correlation functions} 
\subsection{Phase-matched feature of chiral correlation functions in our scheme}
Here, we will show that the nonzero chiral correlation functions of our scheme are phase-matched. The chiral correlation functions~\cite{ayuso2019synthetic} are
\begin{align}\label{hn}
h^{(n)}(\bar{\omega}_0,...,\bar{\omega}_{n-1})=&\sum^{2}_{\{c_i\}=1}\{\bm{F}_{c_{0}}(\bar{\omega}_0)\cdot [\bm{F}_{c_{1}}(\bar{\omega}_1)\times \bm{F}_{c_{2}}(\bar{\omega}_2)]\}[\bm{F}_{c_{3}}(\bar{\omega}_3)\cdot \bm{F}_{c_{4}}(\bar{\omega}_{4})]...\nonumber\\
&\times[\bm{F}_{c_{n-2}}(\bar{\omega}_{n-2})\cdot \bm{F}_{c_{n-1}}(\bar{\omega}_{n-1})], 
\end{align}
where the integer $n$ should be odd and the summation means $\sum^{2}_{\{c_i\}=1}=\sum^{2}_{c_0=1}...\sum^{2}_{c_{n-1}=1}$. The frequencies satisfy 
\begin{align}\label{wn}
\sum^{n-1}_{i=0}\bar{\omega}_{i}=0.
\end{align} 
$\bm{F}_{c}(\bar{\omega})=\bm{F}_{c}^{\ast}(-\bar{\omega})$ is the Fourier transform of the electric field vector at frequency $\bar{\omega}$. Here, we have grouped them by subscripts $c=1,2$. The terms with subscript $c=1$ result from the first sub-beam of our scheme, which is polychrome and linearly polarized. For terms with subscript $c=2$, they result from the second sub-beam of our scheme, which is monochrome and exhibits circular polarization in the plane perpendicular to the polarization of the first sub-beam.

At a fixed position, the overall electric field for a general case of our scheme with $N$-color $c=1$ sub-beam is 
\begin{align}\label{OFE}
\bm{E}(\bm{r},t)=&\sum^{N}_{m=1} \frac{E_{m}}{2}\bm{e}_z[e^{-i(\omega_m t- k_m x+\phi_m)}+c.c.]
+\frac{E_0}{2}[e^{-i(\omega_0 t-k_0 z+\phi_0)}\bm{e}_{\sigma}+c.c.],
\end{align}
where $\sigma(=\pm1)$ denotes the handedness of the circularly polarized field with the polarization vector $\bm{e}_{\sigma}=(\sigma\bm{e}_x+i\bm{e}_y)/\sqrt{2}$, $\phi_m$ and $\phi_0$ are the initial phases, and the amplitudes $E_{m}$ are positive. 
The nonzero Fourier transforms of our electric field vector are 
\begin{align}\label{FTE}
&\bm{F}_{1}(\omega_m)=\bm{F}^{\ast}_{1}(-\omega_m)=\frac{E_{m}}{2}e^{i(k_m x-\phi_m)}\bm{e}_z,~~\bm{F}_{2}(\omega_0)=\bm{F}^{\ast}_{2}(-\omega_0)=\frac{E_0}{2}e^{i(k_0 z-\phi_0)}\bm{e}_{\sigma}.
\end{align} 
It is known that the nonzero terms corresponding to $\bm{e}_{\sigma}$ are 
\begin{align}\label{NZE}
&\bm{e}_{\sigma}\cdot (\bm{e}_z\times\bm{e}^{\ast}_{\sigma})\ne0,~~\bm{e}^{\ast}_{\sigma}\cdot (\bm{e}_z\times\bm{e}_{\sigma})\ne0,~~
\bm{e}_{\sigma}\cdot \bm{e}_{\sigma}^{\ast}\ne 0. 
\end{align}

With Eqs.~(\ref{hn},\ref{FTE},\ref{NZE}), we conclude that  $\bm{F}_{2}({\omega_0})$ and $\bm{F}_{2}(-{\omega_0})$ enter Eq.~(\ref{hn}) of a nonzero $h^{(n)}$ as a pair. Due to the requirement in the frequency [see Eq.~(\ref{wn})], the $c=2$ sub-beam forms an even-photon cyclic process and the $c=1$ sub-beam forms an odd-photon cyclic process. Because each sub-beam propagates collinear, the spatially dependent phases corresponding to each beam cancel each other, yielding spatially uniform $h^{(n)}$ with arbitrary odd integer $n$, i.e., phase-matched.

\subsection{Two-color phase-matched LCLs}
Here, we give another candidate of phase-matched LCL in our scheme, which is composed of two color carrier waves. At a fixed position, the overall electric field is 
\begin{align}\label{OFE2}
\bm{E}(\bm{r},t)=&\sum^{2}_{m=1} \frac{E_{m}}{2}\bm{e}_z[e^{-i(m\omega t- mk x+\phi_m)}+c.c.]
+\frac{E_0}{2}[e^{-i(\omega t-k z+\phi_0)}\bm{e}_{\sigma}+c.c.].
\end{align} 
We note that the overall fields at different positions can be superposed on each other by spatial rotations and temporal translations in the case that the applied fields are approximated as plane waves. This property can be demonstrated analytically because the overall fields can be rearranged by using the rotated direction vectors of electric fields $\tilde{\bm{e}}_{\mu}(\theta)$ ($\mu=X,Y,Z$) with {$\theta=-\sigma k(z-x)$} and the translated time $t^{\prime}=t-kx/\omega$ as
\begin{align}\label{EV}
\bm{E}(\bm{r},t)=\sum_{\mu}E_{\mu}(t^{\prime})\tilde{\bm{e}}_{\mu}(\theta).
\end{align} 
Here, the components of the electric field vector are $E_{X}(t^{\prime})=\sigma E_0\sin(\omega t^{\prime}+\phi_0)/\sqrt{2}$, $E_{Y}(t)=E_0\cos(\omega t^{\prime}+\phi_0)/\sqrt{2}$, and $E_{Z}(t^{\prime})=E_1\cos(\omega t^{\prime}+\phi_1)+E_2\cos(2\omega t^{\prime}+\phi_2)$. Specifically, the rotated direction vectors of the electric fields are $\tilde{\bm{e}}_Z(\theta)=\bm{e}_z$, $\tilde{\bm{e}}_{X}(\theta)=\bm{e}_{x}\cos\theta-\bm{e}_y\sin\theta$, and $\tilde{\bm{e}}_{Y}(\theta)=\bm{e}_{x}\sin\theta+\bm{e}_y\cos\theta$ with the unrotated direction vectors $\bm{e}_x$, $\bm{e}_y$, and $\bm{e}_z$. According to the symmetry-based overlap measure of the degree of (local) chirality~\cite{ayuso2019synthetic}, Eq.\,(\ref{EV}) indicates that the DOCs at different positions in space are identical in the plane-wave limit. For the realistic cases of Gaussian beams as in the main text, the identical feature is replaced with the phase-matched feature for the two-color one of our scheme. 

To further illustrate its phase-matched feature, we give its lowest order of CCFs, which is 
\begin{align}\label{h5}
 h^{(5)}(-2\omega,-\omega,\omega,\omega,\omega)=\frac{1}{32}i \sigma e^{i\phi}E^2_0E^2_1E_2
 \end{align}
with $\phi=\phi_2-2\phi_1$. Here, we can clearly see that the phases $\bm{k}\cdot \bm{r}$ are canceled out, and thus the two-color LCLs of our scheme is phase matched. Although the carrier waves were not simultaneously coupled with chiral molecules in the one-photon (near-)resonance manner, two-color fields can be easily obtained and widely used in the strong-field region.

\subsection{Tricolor locally chiral light}\label{3LCL}
The tricolor LCLs~\cite{kral2003two,ye2018real,leibscher2019principles} are composed of three carrier waves in the three-photon resonance condition. The polarizations of the carrier waves should be well designed~\cite{kral2003two,ye2018real,leibscher2019principles}. Specifically, they can be mutually perpendicular to each other. In this case, the overall field can be written as
\begin{align}
\bm{E}(\bm{r},t)=&\frac{E_{1}}{2}e^{i(\omega_1 t-k_1z+\phi_1)}\bm{e}_x+\frac{E_{2}}{2}e^{i(\omega_2 t-k_2z+\phi_2)}\bm{e}_y+\frac{E_{3}}{2}e^{i(\omega_3 t-k_3 x+\phi_3)}\bm{e}_z+c.c.
\end{align} 
with the three-photon resonance condition
\begin{align}
\omega_1+\omega_2-\omega_3=0.
\end{align}
We note that $\omega_1\neq\omega_2$, otherwise, the overall field is not locally chiral. 

The carrier waves $E_1$ and $E_2$ are in the same sub-beam, labeled as $c=1$. The carrier wave $E_3$ is in the $c=2$ sub-beam. Then, the Fourier transforms are 
\begin{align}
&\bm{F}_1(2\omega)=\frac{E_{3}}{2}e^{i(k_3 x-\phi_3)}\bm{e}_z,~~
\bm{F}_2(\omega_1)=\frac{E_{1}}{2}e^{i(k_1z-\phi_1)}\bm{e}_x,~~
\bm{F}_2(\omega_2)=\frac{E_{2}}{2}e^{i(k_2z-\phi_2)}\bm{e}_y.
\end{align}
The lowest order of the chiral correlation function is 
\begin{align}
h^{(3)}(-2\omega,\omega_1,\omega_2)=\frac{1}{8}E_1E_2E_3e^{i\phi}e^{i[(k_1+k_2)z-k_3 x)]}
\end{align}
with $\phi\equiv\phi_3-\phi_2-\phi_1$. Here, $[(k_1+k_2)z-k_3 x]$ enters the chiral correlation function and thus tricolor LCLs are phase-mismatched.

\subsection{Locally-and-globally chiral light}
The locally-and-globally chiral light (LGCL)~\cite{ayuso2019synthetic} consists of two sub-beams that propagate along $\bm{k}_1$ and $\bm{k}_2$, in the $x-y$ plane, creating angles $\pm \alpha$ with the $y$ axis. The propagation vectors are 
\begin{align}
\bm{k}_1=k\sin(\alpha)\bm{e}_x+k\cos(\alpha)\bm{e}_y,~~
\bm{k}_2=-k\sin(\alpha)\bm{e}_x+k\cos(\alpha)\bm{e}_y.
\end{align}
There are two carrier waves in each beam. The first carrier wave is a linearly polarized $\omega$ field with polarization direction in the $x-y$ plane. The second carrier wave is a $z$-polarized $2\omega$ field. Their electric fields can be written as 
\begin{align}
\bm{E}_{n}(\bm{r},t)=&\frac{E_{0}}{2}[e^{i(\omega t-\bm{k}_n\cdot \bm{r}+\phi^{\omega}_{n})}\bm{e}_n+r_0 
e^{i(2\omega t-2\bm{k}_n\cdot \bm{r}+2\phi^{2\omega}_{n})}\bm{e}_z+c.c.,
\end{align}
where we have 
\begin{align}
&\bm{e}_1=\cos(\alpha) \bm{e}_x-\sin(\alpha)\bm{e}_y,~~
\bm{e}_2=\cos(\alpha)\bm{e}_x+\sin(\alpha)\bm{e}_y. 
\end{align}
The Fourier transforms are 
\begin{align}
&\bm{F}(\omega)=\bm{F}^{\ast}(-\omega)=\sum_{n=1,2}\frac{E_{0}}{2}e^{i(\bm{k}_n\cdot \bm{r}-\phi^{\omega}_{n})}\bm{e}_n,\nonumber\\
&\bm{F}(2\omega)=\bm{F}^{\ast}(-2\omega)=\sum_{n=1,2}r_0 \frac{E_{0}}{2}
e^{i(2\omega t-2\bm{k}_n\cdot \bm{r}+2\phi^{2\omega}_{n})}\bm{e}_z].
\end{align}
The lowest order of the chiral correlation function is 
\begin{align}\label{h51}
h^{(5)}(-2\omega,-\omega,\omega,\omega,\omega)
=&\bm{F}^{\ast}(2\omega)\cdot[\bm{F}^{\ast}(\omega)\times\bm{F}(\omega)][\bm{F}(\omega)\cdot\bm{F}(\omega)]\\
=&-(1/4) i E_0^5 r_0 \{\cos(2 \alpha)- \cos[2 k x \sin(\alpha)]\}\sin(2 \alpha)\sin[2 k x \sin(\alpha)]^2,\nonumber
\end{align}
where $kx$ enters the chiral correlation function and thus the LGCLs are phase-mismatched.  

\section{Enantiospecific electronic state transfer}

\subsection{An example of methyloxirane: four-color phase-matched LCL}

\begin{figure}[htp]
	\centering
	\includegraphics[width=0.6\columnwidth]{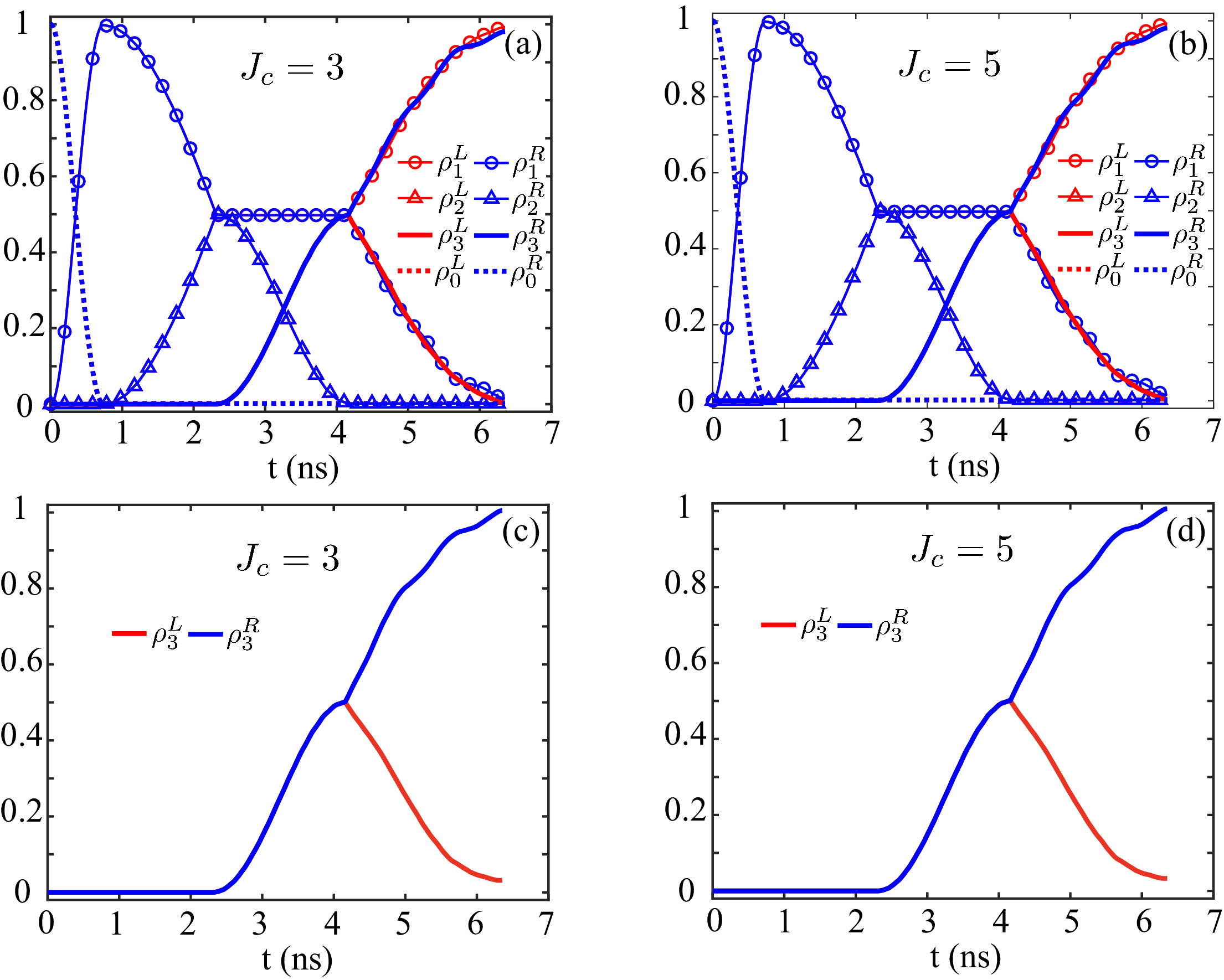}\\
	\caption{The evolution of the population in each electronic state for the initial state of $|0\rangle|0_{0,0,0}\rangle$. (a) $J_c=3$ and (b) $J_c=5$, which clearly show the convergence of the numerical simulation at $J_c=3$. The evolution of the population in the electronic state $|3\rangle$ for the initial state of Eq.\,(\ref{TM}) at $T_{\mathrm{rot}}=1$\,K: (c) $J_c=3$ and (d) $J_c=5$.}\label{S2}
\end{figure}

The working electronic states $|0\rangle$, $|1\rangle$, $|2\rangle$, and $|3\rangle$ are chosen as the electronic ground state, the $3p_y$-state, the $3p_x$-state, and the $3d_{z^2-x^2}$-state. The carrier waves with angular frequencies $\omega_0$, $\omega_1$, $\omega_2$, and $\omega_3$ are chosen to resonantly couple with transitions $|0\rangle|0_{0,0}\rangle\leftrightarrow|1\rangle|1_{0,1}\rangle$, $|1\rangle|1_{0,1}\rangle\leftrightarrow|2\rangle|2_{1,1}\rangle$, $|2\rangle|2_{1,1}\rangle\leftrightarrow|3\rangle|2_{0,2}\rangle$, and $|1\rangle|1_{0,1}\rangle\leftrightarrow|3\rangle|2_{0,2}\rangle$. The wavelengths are $\lambda_0\simeq162.2633$\,nm, $\lambda_1\simeq6241.4404$\,nm, $\lambda_2\simeq1541.6517$\,nm, and $\lambda_3\simeq2047.3531$\,nm. These parameters are same as those in the main text.

Firstly, we would like to show that in the case of the initial state $|0\rangle|0_{0,0,0}\rangle$ the cutoff of $J$ is obtained at $J_c=3$. In Fig.\,\ref{S2}\,(a) and Fig.\,\ref{S2}\,(b), we compare the population evolutions for the cases of $J_c=3$ and $J_c=5$, which clearly show that the results of the two cases agree with each other. Moreover, we note that the enantiomeric excesses of the two cases are $\varepsilon=99.27\%$ and $\varepsilon=99.26\%$, respectively, which further confirm that for the case of the initial state $|0\rangle|0_{0,0,0}\rangle$ the cutoff of $J$ is obtained at $J_c=3$. 

Secondly, we would like to explore the effect of the thermal distributions in the excited rotational states on the enantiomeric excess. We take the typical rotational temperature of supersonic molecular beam ${T}_{\mathrm{rot}}=1$\,K. We take the relative Boltzmann distribution of rotational state in the electronic ground state with respect to the ground state $|0\rangle|0_{0,0,0}\rangle$
\begin{align}
\rho(\mathcal{R})=\exp\left(\frac{-E_{\mathcal{R}}}{k_b T_{\mathrm{rot}}}\right),
\end{align} 
where $E_{\mathcal{R}}$ is the energy of the rotational eigenstate $|\mathcal{R}\rangle=|J_{K_a,K_c,M}\rangle$ and $k_b$ is the Boltzmann constant. Then, the initial state is given as 
\begin{align}\label{TM}
\hat{\rho}(T_{\mathrm{rot}})=\sum_{\mathcal{R}}\rho(\mathcal{R})|\mathcal{R}\rangle\langle \mathcal{R}|\otimes|0\rangle\langle 0|.
\end{align}
We note that the excited electronic states are well separated from the ground electronic state, such that their relative Boltzmann distributions are assumed $zero$.
The results for $J_c=3$ and $J_c=5$ are given in Fig.\,\ref{S2}\,(c) and Fig.\,\ref{S2}\,(d), respectively. The corresponding enantiomeric excesses are $-93.63\%$ and $-93.83\%$. We note that although there is a considerable portion of the population in rotational states with $J>5$, this portion of molecules is almost undisturbed because corresponding transitions are off-resonantly coupled and the applied fields are weak. The global enantiomeric excess for the initial state of Eq.\,(\ref{TM}) at $T_{\mathrm{rot}}=1$\,K with $J_c=3$ is $\varepsilon_{\mathrm{global}}\simeq-91\%$. 

\subsection{Enantiosepcific state transfer by using tricolor LCLs}

\begin{figure}[hpt]
	\centering
	\includegraphics[width=0.6\columnwidth]{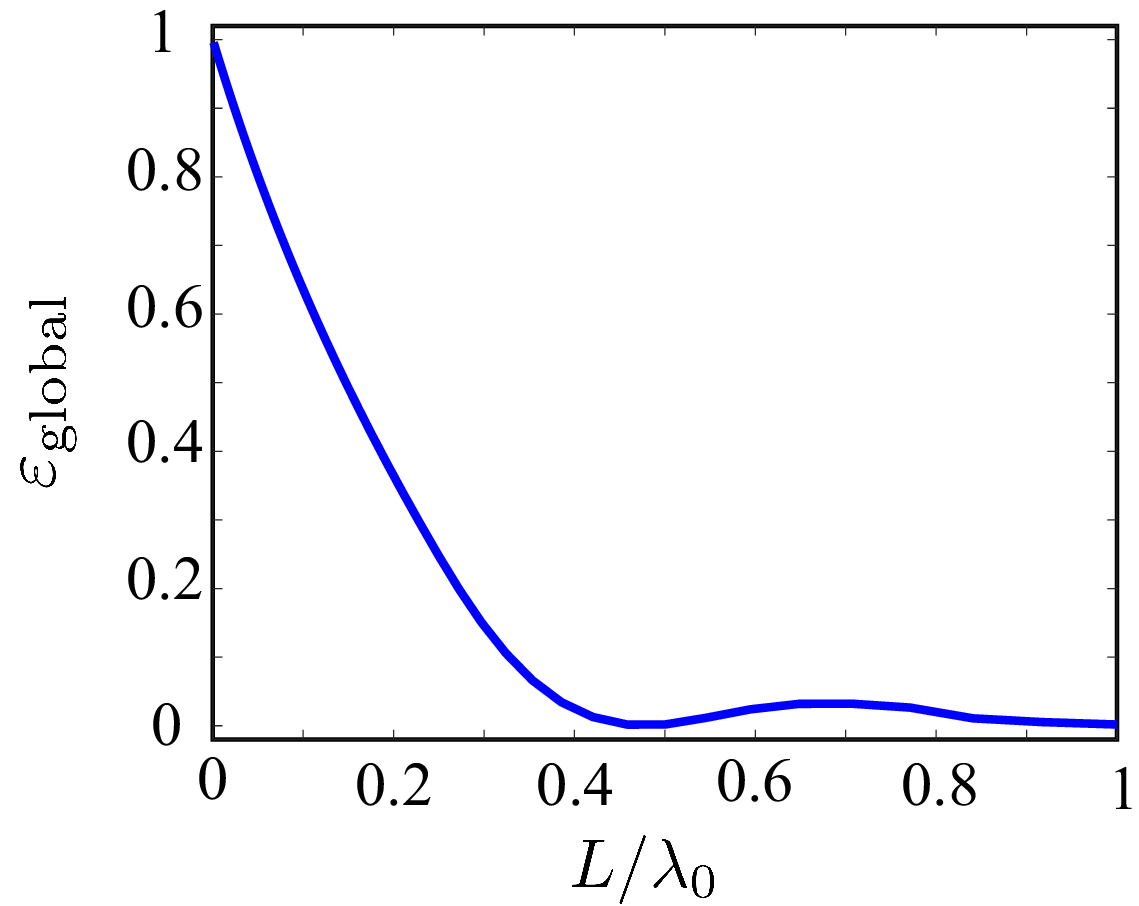}\\
	\caption{Global enantiomeric excess as the function of $L$ for tricolor LCLs. We assume the carrier waves are plane waves. The sample is assumed a $L\times L\times L$ cube. The carrier wave with the smallest absolute values of $\bm{k}$ propagates in the direction vertical to one face of the cube. }\label{S3}
\end{figure}

In order to show the global efficiency of enantiospecfic state transfer of tricolor LCLs, we use the cyclic three-level Hamiltonian~\cite{ye2018real} 
\begin{align}\label{HM}
    \hat{\mathcal{H}}_{Q}=&\frac{1}{2}{\hbar}[\Omega^{Q}_{10}|\tilde{1}\rangle\langle \tilde0|
   +\Omega^{Q}_{20}|\tilde2\rangle\langle \tilde0|+\Omega^{Q}_{21}e^{\delta \bm{k}\cdot \bm{r}}|\tilde2\rangle\langle \tilde1|+h.c.],
\end{align}
with $\delta\bm{k}=\bm{k}_{20}-\bm{k}_{21}-\bm{k}_{10}$ and $\Omega^{L}_{jl}=-\Omega^{R}_{jl}$. Due to the specific requirement of the polarization direction of the three carrier waves~\cite{ye2018real} (or see Sec.\,\ref{3LCL}), the smallest $|\delta\bm{k}|$ is 
\begin{align}
|\delta\bm{k}|=\sqrt{2}\min(|\bm{k}_{10}|,|\bm{k}_{21}|,|\bm{k}_{20}|)=\frac{2\sqrt{2}\pi}{\lambda_0}. 
\end{align}
This is obtained in the case that the carrier wave with the smallest absolute values of $\bm{k}$ propagates vertically to the propagating direction of the other two carrier waves and the other two carrier waves propagate in the same direction.

For simplicity, we assume the carrier waves are plane waves. The sample is assumed a $L\times L\times L$ cube. The carrier wave with the smallest absolute values of $\bm{k}$ propagates in the direction vertical to one face of the cube. Then, the global enantiomeric excess as the function of $L$ is given in Fig.\,\ref{S3}. The results show that the global enantiomeric excesses decrease quickly with the sample size increase. To obtain high-efficiency global control ($|
\varepsilon_{\mathrm{global}}|\ge90\%$), the size should be $L\le 0.02\lambda_0$. Therefore, the tricolor LCLs meet a grave problem when the wavelengths are short, such as the cases of electronic transitions, vibrionic transitions, and vibrational transitions. In these cases, the sample with $L\le 0.02\lambda_0$ will be extremely small, which is difficult to obtain.

\bibliography{ycref}
